# Optoelectronic and Electronic properties of cubic bulk and 001 surface of CsPbBr$_3$

H. Joshi[1], R. K. Thapa[2], Lalrinthara Pachuau[3], Lalmuanpuia Vanchhawng[3] and D. P. Rai[3*]

[1]*Department of Physics, St. Joseph's College, Darjeeling India*

[2]*Retired Professor, Department of Physics, Mizoram University, Aizawl, India*

[5]*Physical Sciences Research Center (PSRC), Deparment of Physics, Pachhunga University College, Mizoram University, Aizawl-796001, India*

**Abstract:** First Principles calculations are performed on the cubic bulk and its (001) surface of CsPbBr$_3$ to reveal its electronic and optoelectronic properties. Various calculation schemes like GGA, GGA+SOC and the highly accurate modified Becke Johnson potential (mBJ) are applied. The electronic energy band gap calculations obtained from mBJ shows close approximity with the available experimental results. The cubic bulk phase is investigated for its mechanical stability and is found to bypass other structural phases. The optical absorption of both the bulk and the surface surpass non-lead based cubic halides and zinc based organic-inorganic hybrids.

## Introduction

Inorganic lead halide perovskite materials have re-emerged in solid-state solar cell research due to their improved carrier mobility and high stability, compared to the hybrid organic-inorganic analogue [1]. The solid-state inorganic single perovskite material has a general chemical formula AMX$_3$, where A is a cation (Li, Na, K, Rb, Cs, etc.), M is a second cation (typically Pb/Sn) and X is a halide (F, Cl, Br, I) [2-4]. Several other oxide based single perovskite (AMO3) has been studied from ab-initio theory [5–8] and experiment [9, 10] but their high temperature stability and surface oxidation due to exposure to environment is still questionable. The hybrid organic-inorganic counterpart, such as CH$_3$NH$_3$PbBr$_3$, offers high solar efficiency but has limited stability when retained within the crystal lattice due to weak chemical bonding of the corresponding organic cations [11]. The inorganic components in turn provide enhanced stability of the crystal lattice and their mixed anion and cation compositions allow better tunability. Stability plays a crucial role for practical application perspective as well as for various methods used in tailoring the desired property of the material. Almost, all-inorganic lead halide perovskites such as CsPbBr$_3$, CsPbI$_3$, and their doped alloy CsPbBr$_x$I$_{3-x}$ have better stability while their optoelectronic properties are in line with organic-inorganic hybrids [12]. Naturally occurring CsPbBr$_3$ shows orthorhombic crystal structure and is stable up to temperatures below 88º C [13]. Distortion in the orthorhombic symmetry is observed for higher temperatures exceeding 88º C and is converted to tetragonal structure which is stable up to 130º C. For temperatures exceeding 130º C, the cubic symmetry subsequently gains possession over the orthorhombic and tetragonal structures [14]. However, the trichloride lead halide perovskites shows stability at lower ranges of temperatures, where the orthorhombic symmetry transforms to tetragonal at 42º C and the cubic phase overtakes the structural symmetry from 47º C temperature onwards [15]. Owing to the temperature ranges, the tribromide structure possesses higher temperature resistance in comparison to tribromide or trichhloride ones.

In this work, we provide an analysis on the photovoltaic and optoelectronic applications of cubic CsPbBr$_3$ perovskite and its (001) surface employing density functional theory. We employ the most accurate modified Becke Johnson semi-local potential (mBJ) [16] to compare the calculated energy band gaps with the available experimental results. The different electronic states responsible for semiconducting behavior and the optical absorption coefficient in the visible and lower ultra violet regions, important for optoelectronic and photovoltaic applications are discussed in detail. It is observed that Cs atoms do not contribute in energy band formation and dedicate itself solely to high structural stability of cubic $CsPbBr_3$. The electronic structure and the optical response of (001) surface is the first ever reported to the best of our knowledge. The optical absorption in the visible and the lower ultraviolet region of the studied bulk and surface bypass non-lead based cubic halides $CsXBr_3$ (X=Ge, Sn) [3] and the zinc based organic-inorganic hybrids [17].

**Computational Details**

For computation of both the cubic bulk and surface of $CsPbBr_3$ we have used Kohn-Sham density functional theory (KS-DFT) [18, 19] based WIEN2K code [20] which incorporate full potential linearized augmented plane wave (FP-LAPW) basis set. In this technique, the atomic space is sub-divided into two regions namely, the non-overlapping atomic sphere region (Region I), also known as the muffin-tin (MT) sphere region and the interstitial region (Region II), i.e. the space between two spheres. In both the regions, diversed basis sets are applied. In region I, generalize eigen value equations are obtained through diagonalization. In region II, the basis functions are the solutions of kinetic energy dependent Helmholtz spherical equation. The so formed Fourier series corresponding to region II, is the pseudo-wave function. The crystal wave functions are toned to interstitial crystal wave at boundaries to ensure their continuity and differentiability. Potential obtained via charge density through Poisson's equation is utilized to solve radial Schrodinger equation. Overlapping atomic charge density is taken as the density for first iteration. A new solution is obtained by constructing a fresh charge density derived from eigenvectors calculated variationally. This is repeated until self-consistency criterion is fulfilled for which the following parameters were set for calculation- Within the MT sphere, the highest value of angular momentum function ($l_{max}$) for non-spherical charge density and potential, was set to 10. The cut-off parameter was set to $R_{MT} \times K_{max} = 7$, where $K_{max}$ is the highest value of reciprocal lattice vector in the plane wave expansion and $R_{MT}$ is the smallest atomic sphere radii of all atomic spheres. A dense k-mesh of 20×20×20 is considered in the first Brillouin zone for the self-consistent DFT calculation. The self-consistency convergence criteria for energy is set to 0.0001 Ry. To make the electronic properties calculation consistent with the experimental scenario, modified Becke Johnson potential [16] was adopted for convergence over Perdew-Burke-Ernzerhof Generalized Gradient Approximation (PBE-GGA) [21] and LDA [22]. Also, to accurately determine the electronic states involved in electronic energy band, relativistic effects were considered by inclusion of spin-orbit interactions, due to the involvement of heavy atoms like Pb in the crystal lattice. The elastic calculations are obtained employing stress-strain method [23, 24] from the second order derivatives of the polynomial fit of energy vs strain.

**Results**

(a) Structural and Elastic Properties

The cubic structure of $CsPbBr_3$ crystallizes with space group symmetry $Pm\bar{3}m$ (No. 221) and takes the following Wyckoff positions, Cs: 1b (1/2 1/2 1/2); Pd: 1a (0 0 0); Br: 3d (0 1/2 0). The (0 0 1) surface of the cubic $CsPbBr_3$ was obtained from a supercell of the original bulk structure. Since, the first principles calculations are periodic, essentially in all three directions, so for surface calculations, a vacuum layer of thickness 15 Å along the X and Y direction was applied. Thickness of a vacuum layer is crucial to terminate the atomic interactions for 2-D calculations.

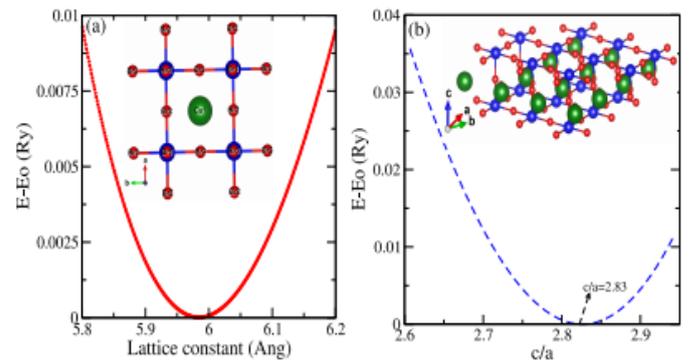

**Figure 1:** Volume optimization curve of **(a)** bulk and **(b)** 001 surface of $CsPbBr_3$ (Atom colour scheme, Cs-green, Br-red, Pb-blue).

The optimized lattice constants were determined by fitting the Energy – Volume curve, obtained through first principles self-consistent calculations, with third order Murnaghan's equation of state [25] as

$$E(V) = E_0 + \frac{9V_0 B_0}{16}[\{(V_0/V)^{2/3} - 1\}^3 B_0' + \{(V_0/V)^{2/3} - 1\}^2 \{6 - 4(V_0/V)^{2/3}\}]$$

where, $E_0$ is the equilibrium energy, $V_0$ the original volume, $V$ the obtained volume change, $B$ is the Bulk modulus and $B'$ its first order pressure derivative. The optimization curves are shown in figure 1 for both bulk and (001) surface. The obtained lattice constants are tabulated in table 1 and are compared with the experimental and available theoretical result.

**Table 1:** Calculated structural parameters, lattice constants (Å), Bulk modulus ($B$) and its pressure derivatives ($B'$) for bulk $CsPbBr_3$.

| This study | | | Others | |
|---|---|---|---|---|
| Å | B | B′ | Å | B |
| 5.98 | 20.07 | | 5.87[a] (Expt.) | 18.14[b] (Theo.) |
| | | | 5.77[b] (Theo.) | 21.45[b] (Theo.) |
| | | | 5.84[c] (Expt.) | 27.00[e] (Expt.) |
| | | | 5.74[d] (Expt.) | |

[a]Ref. [26], [b]Ref. [13], [c]Ref. [27], [d]Ref. [28], [e]Ref. [29]

Also, to illustrate the fundamental parameters like stability and the anisotropy of the cubic phase, various elastic constants has been calculated. The tensor representation of the independent elastic constants $C_{11}$, $C_{12}$ and $C_{44}$ is expressed as

$$C_{ij} = \begin{pmatrix} C_{11} & C_{12} & C_{12} & 0 & 0 & 0 \\ C_{12} & C_{11} & C_{12} & 0 & 0 & 0 \\ C_{12} & C_{12} & C_{11} & 0 & 0 & 0 \\ 0 & 0 & 0 & C_{44} & 0 & 0 \\ 0 & 0 & 0 & 0 & C_{44} & 0 \\ 0 & 0 & 0 & 0 & 0 & C_{44} \end{pmatrix}$$

The three strain tensors ($\delta, \delta, \delta, 0, 0, 0$), ($\delta, \delta, 0, 0, 0, 0$) and ($0, 0, 0, \delta, \delta, \delta$) were used to calculate the elastic constants. The elastic energy change for is expressed respectively as [30]

$$\frac{\Delta E_1}{V} = \frac{3}{2} C_{44} \delta^2 \quad (1)$$

$$\frac{\Delta E_2}{V} = (C_{11} + C_{12}) \delta^2 \quad (2)$$

$$\frac{\Delta E_3}{V} = \frac{3}{2}(C_{11} + C_{12}) \delta^2 \quad (3)$$

The elastic stability criterion expressed by eq. (4) tests the mechanical resistance against deformation.

$$\begin{aligned} & C_{44} > 0, \\ & \frac{C_{11} - C_{12}}{2} > 0, \\ & B > 0, \\ & C_{12} < B < C_{11} \end{aligned} \quad (4)$$

The Young's modulus (E), shear modulus (G), Voigt's shear modulus ($G_V$), Reuss's shear modulus ($G_R$), anisotropy factor ($A_l$, $A_u$, $A_e$), Pugh's ration (B/G), Kleinman parameter ($\xi$), Potter's formula ($Z/Z_0$) and Poisson's ratio ($\sigma$), is then calculated in relation to the obtained elastic constants as

$$E = \frac{9GB}{3B + G} \quad (5)$$

$$G = \frac{G_R + G_V}{2} \quad (6)$$

$$G_V = \frac{C_{11} - C_{12} + 3C_{44}}{5} \quad (7)$$

$$G_R = \frac{5(C_{11} - C_{12})C_{44}}{4C_{44} + 3(C_{11} - C_{12})} \quad (8)$$

$$A_l = \frac{2C_{44}}{C_{11} - C_{12}} \quad (9)$$

$$\sigma = \frac{3B - 2G}{2(3B + G)} \quad (10)$$

$$\xi = \frac{C_{11} + 8C_{12}}{7C_{11} + 2C_{12}} \quad (11)$$

Table 2 shows the elastic constants along with different elastic parameters. It can be seen that $C_{11} > C_{44} > C_{12}$, implying that the material is hard to deform longitudinally than shape deformation. Also, the low values of $C_{44}$ and $C_{12}$ indicates very low resistance of cubic $CsPbBr_3$ to shear deformation. The stability criteria given by eq. (4) is satisfied, establishing mechanical stability of the alloy, which is expected to retain for higher pressures up to 20 GPa. A significant value of Young's modulus ($E$) shows low deformation of the cubic phase against elastic load, compared to the orthorhombic phase with $E$ = 15.8 GPa [31]. Also, its Bulk modulus value surpasses that of orthorhombic phase ($B$ = 15.5 GPa), suggesting better resistance to fractures for cubic structure. The cubic structure possess higher values of elastic moduli than the other phases, making it potential for flexible optoelectronic applications.

The ratio of Bulk modulus to Shear modulus is often associated with brittle or ductile nature of the compound, with a critical value of about 1.75 [32]. A higher value than the critical limit of the ratio corresponds to ductile nature, whereas a lower value indicates brittle behavior. The compound shows value higher than the critical limit and is thus ductile in nature. Another measure for brittleness and ductility is given by the Poisson's ratio ($\sigma$). The values $\sigma < 0.26$ and $\sigma > 0.26$ respectively corresponds to brittle and ductile characteristic. The obtained $\sigma$ value of 0.28 reconfirms the ductile behavior of the material. Kleinman parameter ($\xi$), given by equation 13 for cubic crystal, represents either bending or stretching under applied stress, and takes the value in the range $0 \leq \xi \leq 1$. $\xi = 1$ and $\xi = 0$ respectively indicates minimized bond stretching and minimized bond bending. The isotropic or the anisotropic nature of the compound can be established from Lame's constant $\lambda$ and $\mu$ [33]. The conditions for elastic isotropy are $\lambda = C_{12}$ and $\mu = (C_{11} - C_{12})/2 = C_{44}$. Where $\lambda$ and $\mu$ are calculated as

$$\lambda = \frac{E\sigma}{(1+\sigma)(1-2\sigma)} \quad (12)$$

$$\mu = \frac{E}{2(1+\sigma)} \quad (13)$$

The obtained $\lambda$ and $\mu$ values diverge from isotropic condition of Lame, establishing elastic anisotropy of the cubic phase in CsPbBr$_3$. Another measure of elastic isotropy or anisotropy is given by the value of A. In any crystal, A=0 is an indication of total isotropy, value other than zero indicates elastic anisotropy. The calculated A value of 0.21 reconfirms the anisotropic characteristic. Furthermore, the elastic wave velocities and the Debye temperature calculated from elastic constants [34] are also showcased in table 2.

**Table 2:** Elastic constants ($C_{ij}$) in GPa, Bulk moduli (B) in GPa, Young's modulus (E) in GPa, Shear Moduli (G) in GPa, Anisotropy ($A_l$, $A_u$, $A_e$), B/G, Kleinman parameter ($\xi$), Elastic velocities ($v_l, v_t, v_m$) in m/s and Debye Temperature ($\Theta_D$) in K.

|  | $C_{11}$ | $C_{12}$ | $C_{44}$ | B | E | G | B/G | $\sigma$ | A | $\xi$ | $v_l$ | $v_t$ | $v_m$ | $\Theta_D$ |
|---|---|---|---|---|---|---|---|---|---|---|---|---|---|---|
| Calc. | 50.3 | 4.9 | 5.7 | 20.1 | 49.4 | 10.3 | 1.94 | 0.28 | 0.21 | 0.25 | 1226.4 | 677.4 | 174.5 | 14.9 |
| Ref. [35] | 42.3 | 6.5 | 4.2 | 19.5 | 41.3 | 9.7 | 2.33 | 0.31 | ××× | ××× | ××× | ××× | ××× | ××× |
| Ref. [36] | 41.2 | 5.4 | 4.1 | 17.3 | 39.9 | 7.8 | 2.21 | ××× | ××× | ××× | ××× | ××× | ××× | ××× |

(b) Electronic Properties

The optimized lattice constants obtained from Murnaghan's equation state were used to investigate the density of states (DOS) and electronic band structure characteristics of both bulk and 001 surface of CsPbBr$_3$. The PGE-GGA functional used in the calculation is known to underestimate the experimental energy band gap about 30-40% in case of semiconductors and insulators [37]. However, its wide spread availability in commercial codes and computational speed makes it popular among experimentalists and theorists. The band gap problem arises due to the unreal self-Coulomb repulsion accounted by GGA functional, leading to estimation of energy bands closer to the Fermi level than the actual physical scenario. One of the way to tackle the problem is by reducing the self-Coulomb repulsion error, applying precise Hartee-Fock (HF) exchange but demands high computational cost [38]. The GGA+U method comes handy, which includes calculation of the onsite Coulomb self-interaction potential term (U), to reduce the self-Coulomb repulsion error but is limited to highly correlated electronic states such as 3d or 4f and hence, cannot be applied to our compound CsPbBr$_3$. Therefore, for accurate determination of electronic structure, advanced exchange correlation (XC) approximations like hybrid DFT functionals, containing fraction of precise HF exchange or many body GW approximations has to be applied. However they can only be applied to small systems or interfaces as they are highly computationally expensive. A way out of this vicious situation is by the usage of Trans and Blaha modified Becke Johnson (TB-mBJ) semilocal approximation, known to describe electronic structures even better than HF and hybrid functional methods in some cases [39] and with very cheap computational expense comparable to GGA. TB-mBJ mimics the exact exchange of HF and hybrid functionals without having to compute it, thereby highly reducing the computational demand retaining high accuracy. Also, the electronic properties determined incorporating spin orbit interactions include relativistic corrections to GGA. The structure optimization of the compound exposes the stability of non-magnetic phase over the spin polarized magnetic case and hence, spin polarized effects are excluded in calculating electronic properties.

The following figure 2 illustrates the density of states obtained using three schemes. The valence region of the DOS is dominated by $p$ state of Br atom whereas $p$ state of Pb atom dominates the conduction region, with small contributions from Br $p$ state atoms. The Cs atoms shows negligible contributions in either of valence or conduction region. The effect of mBJ and GGA+SOC is distinct in the figures. The contribution to the DOS by different electronic state of atoms does not alter with the use of different schemes but however shift in DOS peaks are observed. With GGA functional, three prominent DOS peaks are observed at the valence region for electron energy -3.4 eV, -2.3 eV and -1.5 eV. The prominent DOS features remain unchanged with mBJ, except the three peaks are shifted to lower energy ranges and a sharp increase in number of states per electron energy is observed. The increase in number of state accounts to projection of extra states by the mBJ potential which results from the correction of unphysical self-Coulomb repulsion error involved in GGA. Further, within the GGA+SOC scheme, splitting of DOS peaks are observed in the valence region, which should correspond to increased degenerate flat bands in the band structure in comparison with GGA and mBJ obtained band structure. Also, the diminished amplitude of DOS peaks within GGA+SOC is due to the reduction in electronic state occupation with the inclusion of spin-orbital interaction.

The calculated electron energy band gaps are in close agreement with the available theoretical and experimental data (Table 3). The band structure plotted along high symmetry directions of the first Briollouin zone (IBZ) are presented in figure 3. The observed sharp DOS peaks in the valence region of DOS figure corresponds to flat bands in the band structure along Γ-X, X-M and M-Γ symmetry directions. The band structure is dominated by a direct transition character at R symmetry point, which is one of the signature for efficient optoelectronic applications. The nonoverlapping states between the conduction and the valence region denotes semiconducting characteristics and the energy gaps obtained as the difference in the Conduction Band Minimum (CBM) and Valence Band Maximum (VBM) via GGA, mBJ and GGA+SOC are respectively 1.86 eV, 2.25 eV and 1.52 eV. A closer look on the band plots show a small crossing on the Fermi level ($E_F$) by VBM in case of GGA and GGA+SOC, revealing degenerate p-type semiconducting nature. The same feature remains absent in case of band plot obtained by mBJ scheme, which is due to the shifting of states observed in DOS plot. The obtained band gap using mBJ is closest to the available experimental results [40-42] whereas those obtained from GGA+SOC shows maximum variation. The observed electronic structure characteristic is similar to $CsPbI_3$ [43], where also direct band gap at R symmetry point is observed. Usually cubic perovskite shows observed band structure characteristics [44].

A closer analysis of the DOS figures reveals the VBM and CBM characteristics of the compound. The Cs atom is expected to maintain structural stability only and our observation that it do not contribute to either of VBM or CBM is supported by previous available research [45-48]. The $p$ orbitals of the halogen atom (i.e. Br in this case) constitutes the VBM, with small contributions from $s$ orbitals of Pb. The CBM is mostly of Pd $6p$ character with a fraction of $s$ orbitals from halogen atom Br. The overlapping of $s$ and $p$ orbitals attribute to hybridization scheme for band gap formation. The DOS and band structure plots of (001) surface is shown in figure 4 and 5 respectively. The surface DOS characteristic shows no significant variation when compared to the bulk and the DOS contribution is almost similar to that of bulk. Numerous DOS peaks in the valence region is one of the dominant feature observed in the figure and the shifting of peaks as well as increase or decrease in DOS amplitude with different calculation scheme is in line with the bulk nature.

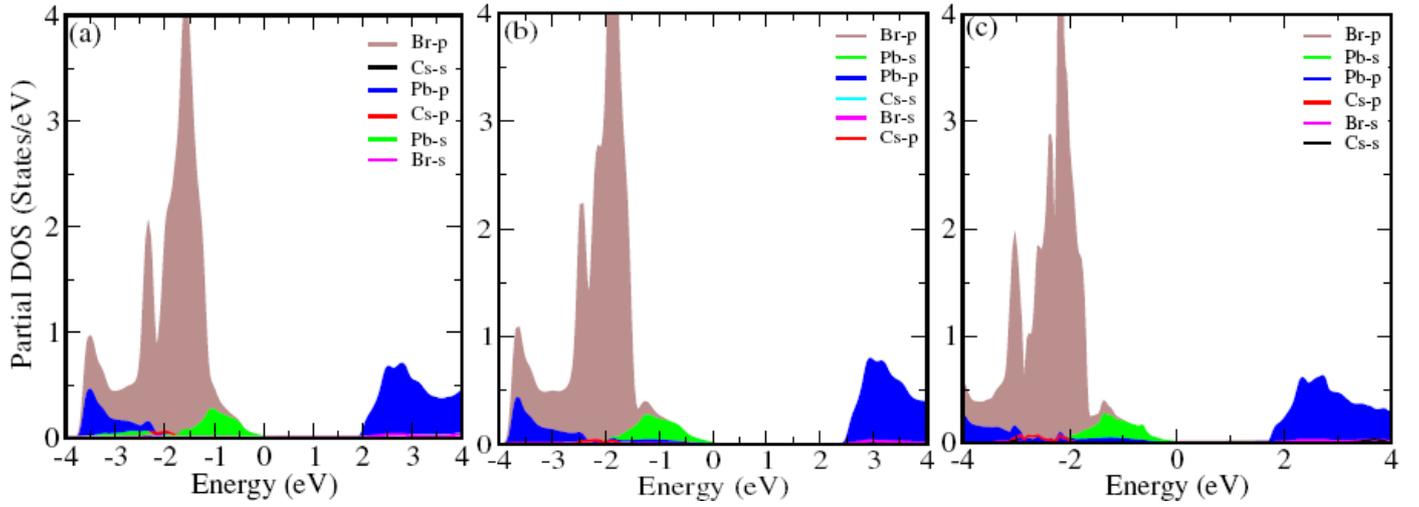

**Figure 2**: Atomic density of states (DOS) of cubic bulk CsPbBr$_3$ calculated from **(a)** GGA **(b)** mBJ and **(c)** GGA+SOC.

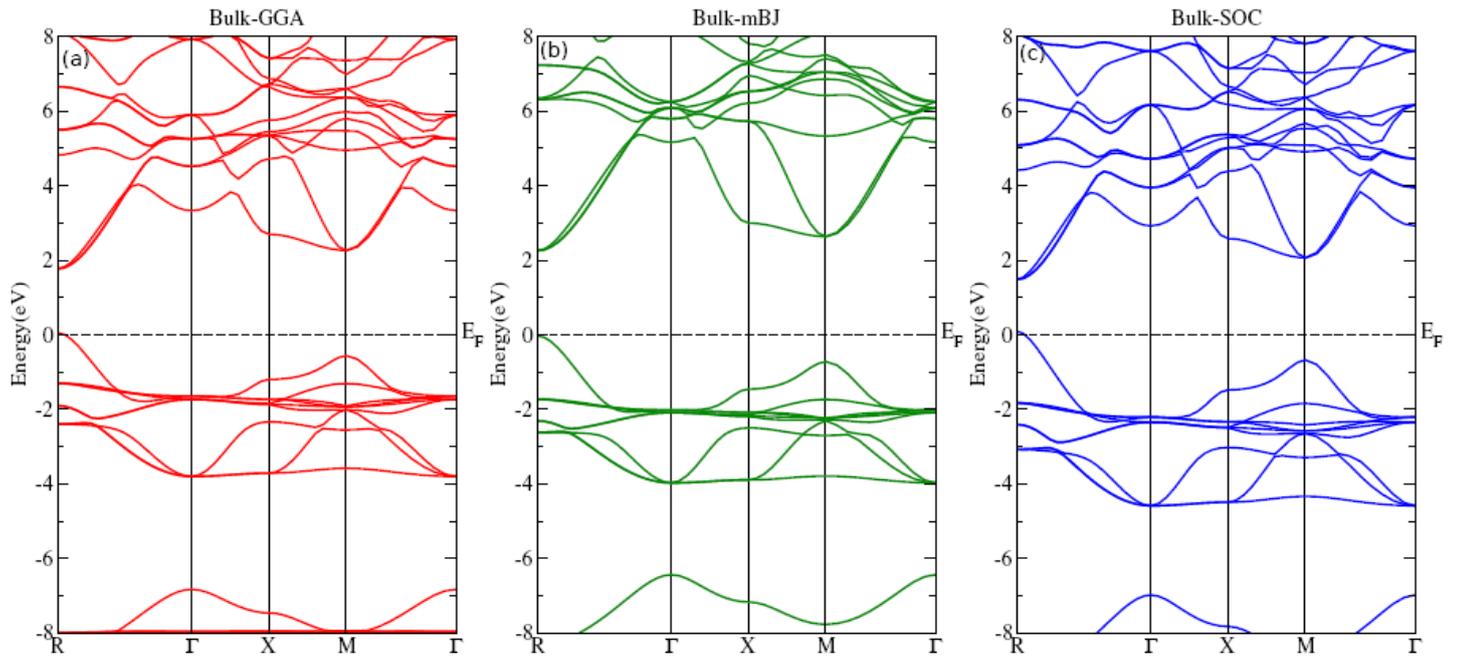

**Figure 3**: Band structure of cubic bulk CsPbBr$_3$ calculated from **(a)** GGA **(b)** mBJ and **(c)** GGA+SOC.

**Table 3:** Calculated energy gap ($E_g$) in eV for cubic bulk CsPbBr$_3$ and its (001) surface using GGA, mBJ and GGA+SOC

| Bulk (Our Calculation) | | | Available Result | | (001) surface | | |
|---|---|---|---|---|---|---|---|
| GGA | mBJ | SOC | Theo. | Expt. | GGA | mBJ | SOC |
| 1.86 | 2.25 | 1.52 | 2.50 (mBJ)[+] | 2.32[*] | 1.76 | 2.36 | 1.79 |
| | | | 2.23 (mBJ)[++] | 2.28[**] | | | |
| | | | 1.05 (SOC)[++] | 2.35[***] | | | |
| | | | 2.34 (GW)[+++] | | | | |

Contrary to the bulk material, VBM and CBM occurs at the M symmetry point but has similar direct transition nature. The Fermi level shows a crossing by the VBM, for band structure obtained with all three schemes indicating p-type degeneracy of the (001) surface. Surprisingly, the energy gap obtained through GGA+SOC gives higher gap value on comparison to GGA. This is mainly because of positioning VBM, which is mostly of Br $4p$ character, at higher energy due to overestimated self-Coulomb repulsion calculated by GGA functional. The band gap obtained from three scheme varies as $E_g$ (mBJ) > $E_g$ (SOC) > $E_g$ (GGA) for the (001) surface.

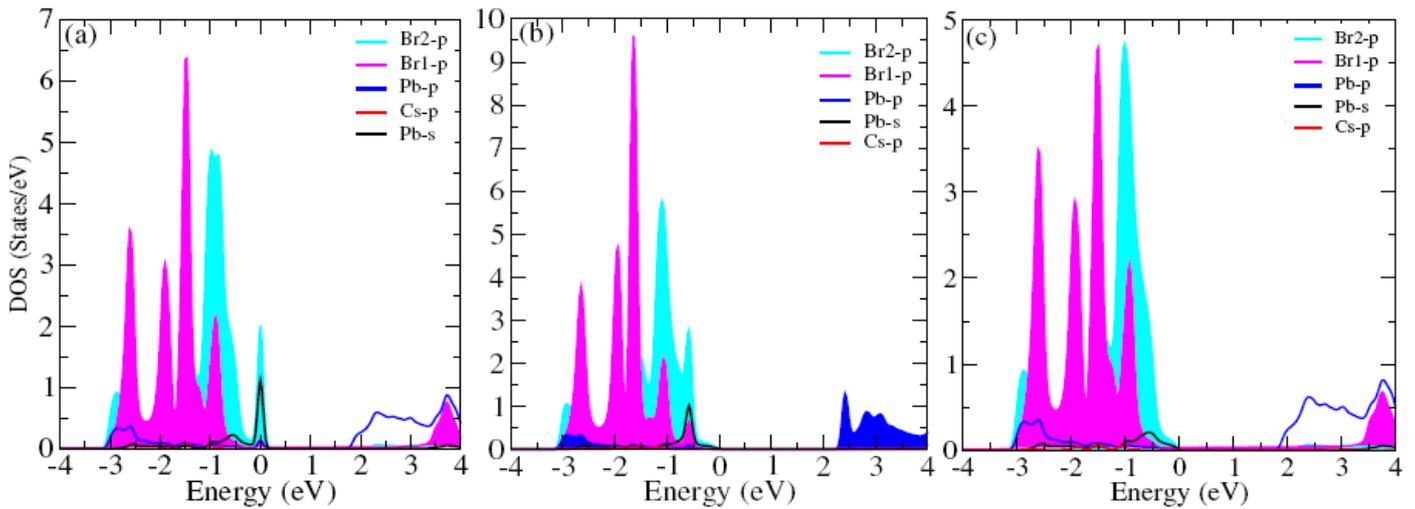

**Figure 4**: Atomic density of states (DOS) of (001) surface of cubic bulk CsPbBr$_3$ calculated from **(a)** GGA **(b)** mBJ and **(c)** GGA+SOC

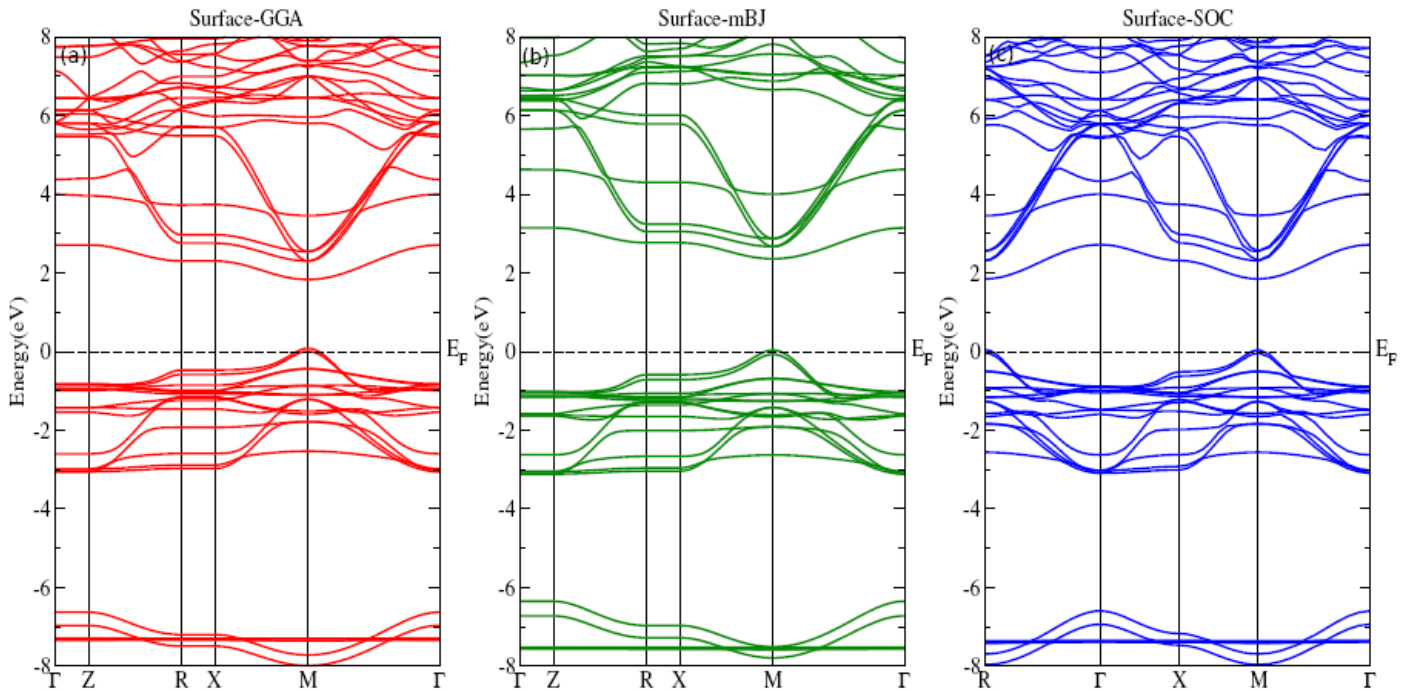

**Figure 5**: Band structure of (001) surface of cubic bulk CsPbBr$_3$ calculated from **(a)** GGA **(b)** mBJ and **(c)** GGA+SOC.

Additionally, the charge density distribution of the (001) surface is investigated to understand the chemical bonding schemes and the electron correlation effects (Figure 6). The localized charge distribution of the valence charge density is a sign of ionic bonding between the electronegative Br⁻ and electropositive $Pb^{2+}$, $Cs^+$, atoms. The isolines in the contour plot indicates covalent nature of bonding. Thus, a mixture of ionic-covalent bond exists in the (001) surface of the cubic $CsPbBr_3$ compound. The covalent character is more prominent for Br-Pb atom bonds, where the isolines can be seen to be shared between the two atoms in the contour plot. The ionic character dominates the (001) surface of the compound as Cs-Cs, Cs-Pb and Cs-Br atom bonds constitutes the ionic character.

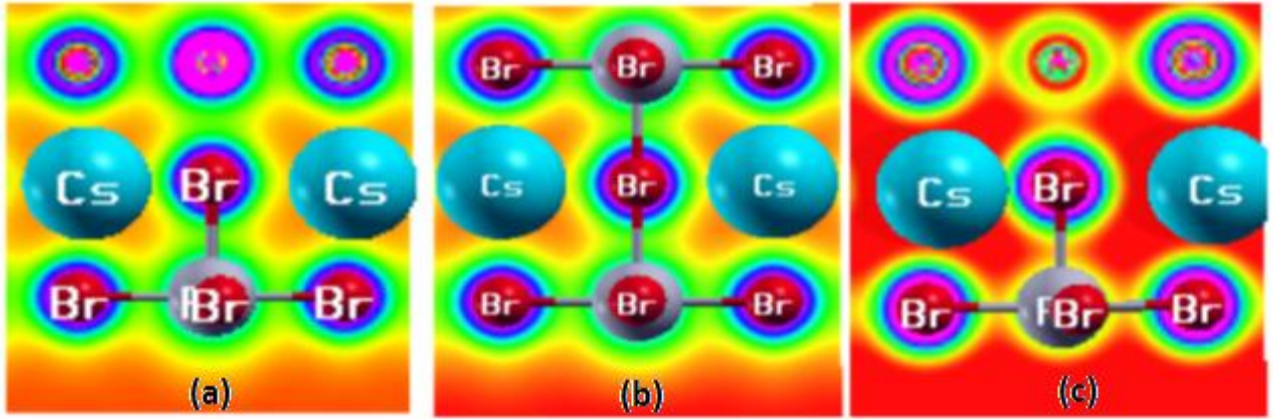

**Figure 6:** Electron density of 001 surface of CrPbBr3 calculated from **(a)** GGA **(b)** mBJ and **(c)** GGA+SOC.

### (c) Optical Properties

A profound knowledge of optical properties is necessary to understand the possibility of a material for photovoltaic and optoelectronic applications. The potentiality of a material for the proposed application is interpreted by examining the electronic transitions to available conduction states from filled valence states, which results from the response of material to incoming electromagnetic radiation. Owing to the semiconducting nature of the material, the inter band transition across valence and conduction states would be dominant over intra band transitions, which is more pronounced in metals. The inter band transition represents all possible transitions between the valence and the conduction states but the direct energy band gap observed for the material designate minimal contributions from shallow lying energy regions in the valence or the conduction band to the optical properties. The frequency dependent complex dielectric function is a major response function from which all other optical constants are obtained and is expressed in real and imaginary part as

$$\varepsilon(\omega)=\varepsilon_1(\omega)+i\varepsilon_2(\omega)$$

The real $\varepsilon_1(\omega)$ and the imaginary $\varepsilon_2(\omega)$ are interrelated and therefore only requires the calculation of imaginary component directly from DFT. $\varepsilon_2(\omega)$ is obtained as the momentum matrices between the occupied and unoccupied electronic eigenstates given by the relation

$$\varepsilon_2(\omega)=\frac{\hbar^2 e^2}{\pi m^2 \omega^2}\sum_{nn'}\int_k d^3k\,|\langle \vec{k}\,n\,|\,\vec{p}\,|\,\vec{k}\,n'\rangle|^2 \times \left[1-f(\vec{k}\,n)\right]\delta(E_{\vec{k}n}-E_{\vec{k}n'}-\hbar\omega) \quad (14)$$

where $\vec{p}$ is the momentum operator, $E_{\vec{k}n}$ is the eigenvalue of the eigen function $|\vec{k}\,n\rangle$ and $f(\vec{k}\,n)$ is the Fermi distribution function. From Kramers-Kronig relation, the real part of the dielectric function is obtained as

$$\varepsilon_1(\omega)=1+\left(\frac{2}{\pi}\right)\int_0^\infty \frac{\varepsilon_2(\omega')\omega'd\omega'}{\omega'^2-\omega^2} \quad (15)$$

The reflectivity R(ω), absorption coefficient α(ω), refractive index n(ω) and other optical constants are then directly obtained from the dielectric functions and are expressed as [51],

$$\alpha(\omega)=\omega\sqrt{2}\left[\sqrt{\varepsilon_1(\omega)^2+\varepsilon_2(\omega)^2}-\varepsilon_1(\omega)\right]^{1/2} \quad (16)$$

$$R(\omega) = \left| \frac{\sqrt{\varepsilon(\omega)} - 1}{\sqrt{\varepsilon(\omega)} + 1} \right|^2 \qquad (17)$$

The following figure 6 and 7 shows the various optical constants representing the optical response of the material to external radiation. The edge values extracted from the imaginary dielectric constants corresponds to the optical gap of the material and are in reasonable agreement with the energy gap values obtained from band structure. Further $\varepsilon_2(\omega)$ represents the absorptive power and is directly related to the band structure. The absorption coefficient $\alpha(\omega)$ quantifies the light penetration within the materials before decaying completely. The semiconductor absorption is related with the photons interacting with electrons, because only those photons can excite electrons which have energy exceeding band gap, otherwise transmission occurs. Fig. 6b and 7b shows absorption increasing above the threshold limit (band gap) and approaching to a peak value due to the incoming photons causing electronic transitions. The instant alteration of the imaginary dielectric function at around 1.7 eV for GGA functional (Figure 7) represents the first direct optical transition and is the optical gap of the compound. The optical gaps are in close relation to the observed values of the band gap, signifying similar types of energy band contribution in the electronic and optical transition. The value of the optical band gap varies similarly with GGA, mBJ and GGA+SOC as observed in the electronic band structure. The (001) surface of the compound shows optical gaps close to the electronic gaps as well, however, the GGA+SOC scheme shows no significant variation than GGA, except slight difference in peak positioning.

The optical absorption (Figure 7b and 8b) in the visible and lower UV region shows significant values. However the material shows best absorption in the UV region. Unfortunately, the absorption value decreases for (001) surface in the visible as well as UV region. The reflectivity (Figure 7c and 8c) plot is consistent with the observed absorption coefficient but the percentage reflectivity at zero frequency $R(0)$ is found to be functional sensitive varying as $R(0)_{mBJ} < R(0)_{GGA} < R(0)_{SOC}$. For the (001) surface, $R(0)$ values are similar for GGA and GGA+SOC and lowest for calculations from mBJ.

The refractive index spectrum closely resembles the $\varepsilon_1(\omega)$, with high peak contributions in the visible region. The values of various optical parameters at zero frequency are listed in table 4.

**Table 4:** Optical gap Zero frequency value of real dielectric constant $\varepsilon_1(0)$, reflectivity R(0), and refractive index n(0) calculated from GGA, mBJ and SOC.

|       | $\varepsilon_1(0)$ | Optical Gap | R(0) %     | n(0)       |
|-------|--------------------|-------------|------------|------------|
| Bulk  | 4.3 (GGA)          | 1.7 (GGA)   | 12 (GGA)   | 2.07 (GGA) |
|       | 3.2 (mBJ)          | 2.5 (mBJ)   | 7.5 (mBJ)  | 1.81 (mBJ) |
|       | 5.2 (SOC)          | 0.7 (SOC)   | 15 (SOC)   | 2.32 (SOC) |
| (001) | 3.4 (GGA)          | 1.5 (GGA)   | 4.2 (GGA)  | 1.57 (GGA) |
|       | 2.2 (mBJ)          | 0.9 (mBJ)   | 3.7 (mBJ)  | 1.45 (mBJ) |
|       | 3.4 (SOC)          | 1.5 (SOC)   | 4.2 (SOC)  | 1.57 (SOC) |

### (d) Conclusion

We investigate the electronic and optical properties of bulk and (001) surface of cubic $CsPbBr_3$ compound. The GGA, mBJ and GGA+SOC schemes has been employed to study the electronic and optical properties. The elastic properties of the bulk structure reveal better mechanical stability in comparison to the other structure class of the compound. The ductile nature of the cubic class is of importance for industrial application. The calculated electronic band gaps are in close relation to the available experimental results. The electronic energy band gap calculated using SOC shows maximum variance from the experimental results whereas that calculated from mBJ has the least variance. The DOS plots reveal various electronic states involved in the hybridization scheme for gap formation. The optical results shows the potentiality of the studied compound and its surface for optoelectronic and photovoltaic applications.

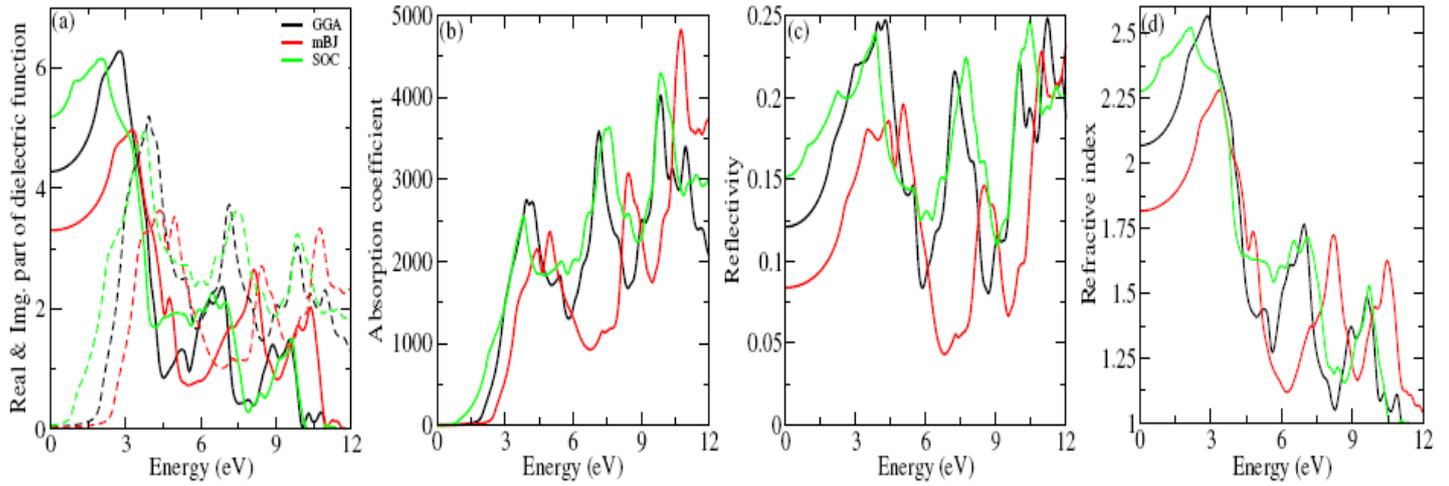

**Figure 6:** **(a)** Real and Imaginary part of dielectric function, **(b)** Absorption coefficient, **(c)** Reflectivity and **(d)** Refractive index of cubic bulk $CrPbBr_3$ calculated from GGA, mBJ and SOC.

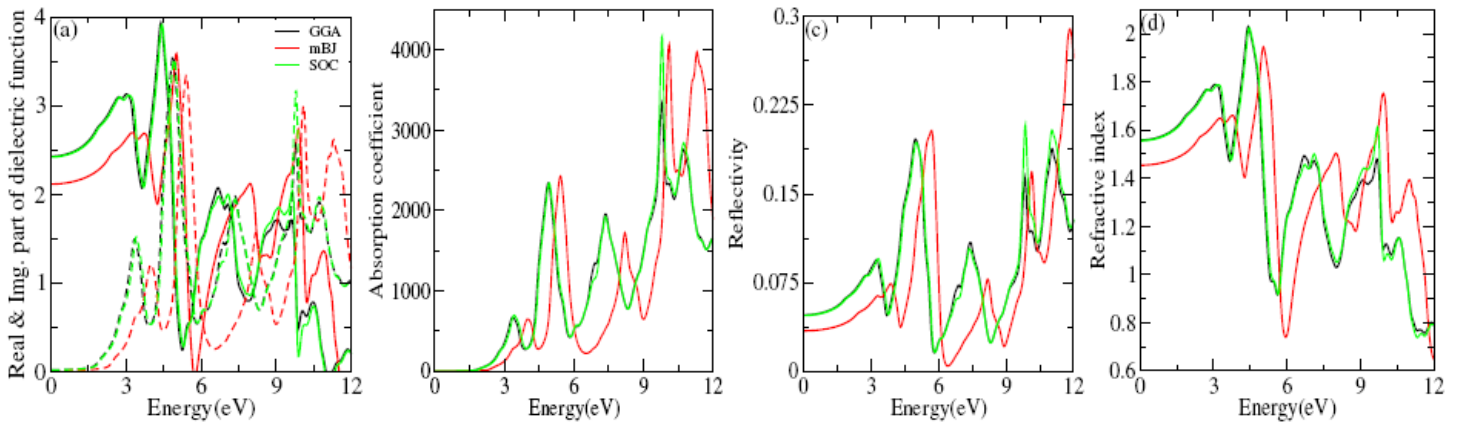

**Figure 7:** **(a)** Real and Imaginary part of dielectric function, **(b)** Absorption coefficient, **(c)** Reflectivity and **(d)** Refractive index of the (001) surface of the cubic bulk $CrPbBr_3$ calculated from GGA, mBJ and SOC.